\documentclass[draftclsnofoot,onecolumn]{IEEEtran} 	
\usepackage{amsmath,amssymb}
\usepackage[mathscr]{eucal}
\usepackage{graphics,graphicx,multicol}
\usepackage{epsfig}
\usepackage{enumerate}
\usepackage{subfigure}
\usepackage{algorithmic}
\usepackage{algorithm}
\usepackage{morefloats}
\usepackage{enumerate}
\usepackage{subfigure}
\usepackage{multirow}
\usepackage{array}
\usepackage{cite}
\usepackage{courier}

\newtheorem{rem}{Remark}[section]

\numberwithin{equation}{section}
\begin{document}

\title{Real-Time Rate-Distortion Optimized Streaming of Wireless Video}

\author{Ahmed Abdelhadi, Andreas Gerstlauer and Sriram Vishwanath \\
}

\maketitle

\begin{abstract}
Mobile cyberphysical systems have received considerable attention over the last decade, as communication, computing and control come  together on a common platform. Understanding the complex  interactions that govern the behavior of large complex cyberphysical systems is not an easy task. The goal of this paper is to address  this challenge in the particular context of multimedia delivery over  an autonomous aerial vehicle (AAV) network. Bandwidth requirements   and stringent delay constraints of real-time video streaming, paired   with limitations on computational complexity and power consumptions  imposed by the underlying implementation platform, make cross-layer and cross-domain co-design approaches a necessity. In this paper, we  propose a novel, low-complexity rate-distortion optimized (RDO)  algorithms specifically targeted at video streaming over mobile embedded networks.  We test the performance of our RDO algorithms using a network of AAVs both in simulation and  implementation.
\end{abstract}


\section{Introduction}
\label{sec:intro}

\subsection{Motivation}\label{sec:motivation}
An explosion of interest in multimedia systems in the last decade has resulted in the need for developing efficient protocols for the delivery of rich content (such as video) across a wireless mobile network. Such media is delay sensitive while being both computationally and bandwidth intensive, and it poses considerable challenges in guaranteeing its reliable delivery.  The main features that make real-time packetized media delivery particularly challenging are \cite{Agrawalrobustdemanddriven}:

\begin{enumerate}
\item \textit{High data rate:} media (especially video) requires  high data rate even when the physical, wireless link between nodes is rapidly changing over time. 
\item \textit{Real-time constraints:} there is a time-to-live (TTL) associated with each packet, so a packet received after its TTL expires is lost. In addition, in live streaming, the video packets has to be delivered within a limited transmitter/camera to receiver/display latency. Otherwise, they will be dropped/lost.
\item \textit{Dependencies between frames:} MPEG4-H.264, MPEG2 and other compressed video  formats are characterized by inter-frame dependency. This has two effects: first, if some frames are not successfully received this will lead to dropping of other successfully received frames because of their interdependency. Second, the number of interdependent frames determines the extent of compression in the video. 
\end{enumerate}
Rich multimedia delivery is increasingly an integral component specifically in many embedded and cyberphysical system applications, which creates additional implementation challenges. Such systems often have to operate in tightly constrained environments that severly limit the available computational performance or the amount of power that can be consumed. Given a vast array of possible embedded and cyberphysical system implementation options and parameters, analyzing and designing protocols and application algorithms\footnote{The word "Algorithm" comes from the name Al-Khwārizmī (c. 780-850), a Muslim mathematician, astronomer, geographer and a scholar in the House of Wisdom in Baghdad. He wrote an algorithm for distributing the inheritance of the deceased to his relatives according to the rules of \textit{Quran}.} for them is a considerably daunting task that invariably requires integrated, cross-layer and cross-domain co-design approaches.

In this chapter, we target the analysis and co-design of multimedia delivery over a particular cyberphysical system consisting of a network of autonomous aerial vehicles (AAVs). Such systems are of considerable interest across multiple civilian and military applications, including search and rescue, perimeter monitoring and object tracking. A network of AAVs poses a vast array of challenges - mobility, tracking and collision avoidance are essential for the physical operation of each AAV, while coordination and teaming critical for the network to carry out the task at hand. Central to all of these challenges is the ability to exchange high bandwidth delay sensitive data between the nodes in the network as reliably and efficiently as possible.

Our ultimate goal is to develop algorithms that exploit the structure of multimedia to deliver them efficiently and reliably over an AAV network, and test them in a real-world setting using a testbed. We have developed our own low-complexity rate-distortion optimized (RDO) streaming algorithms, and show that they outperform other mechanisms in the context of Horus, a custom built AAV testbed \cite{horus:website}. In a first step, we have developed software simulations for the mobility and channel models between AAVs, and we tested both existing and our proposed RDO video streaming techniques using these simulation models.  Results show that optimized streaming can result in much more reliable and efficient video delivery than traditional protocols, in variants both with or without feedback. In the second step, we have implemented the system in realistic settings and tested our proposed RDO protocol and other protocols for different video compression. We used both temporal and spatial distortion measures to 
select the most reliable and efficient protocol.


\subsection{Related Work}
\label{sec:related}

There is considerable existing literature on developing protocols for efficient data delivery over wireless networks. A majority of this literature tends to focus on sensor networks designed for static settings, where nodes
sense physical quantities that undergo a gradual change over time, e.g. temperature. For these applications, only a low data rate is required. Increasingly, sensor networks research incorporates dynamic network topologies as well. In \cite{Juang02energy-efficientcomputing, Liu04implementingsoftware, Zhang04hardwaredesign}, the authors conduct an experimental analysis on a dynamic sensor network where
nodes move in a large area gathering data and then sending the collected data when near an access point. In our network, the nodes move in a prespecified pattern and gather media, e.g. video signals, and
communicate them through the wireless network to an access point in another network. Note that our network is dynamically changing rapidly and intended to support a much higher data rate than conventional
sensor networks.

Simultaneously, there is a growing body of work on media compression and streaming, both over wired and wireless networks. One of these research efforts is presented in \cite{Chou:05}, where the authors
address the problem of streaming packetized media over a lossy network in a rate-distortion optimized way. In \cite{Chou:05}, simulation results show that systems based on rate-distortion optimization (RDO) algorithms have steady-state gains of more than 2-6 dB compared to systems that are not rate-distortion optimized. In this work, a simplified simulation model approximating real-world conditions is assumed, which is only a first step in measuring the performance and expected improvement an algorithm has over existing implementations. For wide-spread system deployment and evaluation of achievable gains of any algorithm, experiments and validations must be carried out in a realistic setting. Towards this goal, we model and deploy RDO implementations in the context of an actual AAV testbed, where realistic simulations are a first step followed by running physical experiments in the field. Furthermore, the original RDO algorithm presented in \cite{Chou:05} is based on ideal assumptions, e.g.\ in terms of its 
implementability. We instead propose modified RDO versions that can be efficiently realized with little to no overhead as part of standard network stacks on restricted embedded platforms.


\subsection{Our Contributions}
Our contributions in this chapter are summarized as:
\begin{itemize}
\item A new low-complexity RDO algorithm \cite{Ahmed3} using default Media Access Control Layer (MAC-L) ACKs, which is validated through real-world simulations and shown to outperform ACKed and not ACKed transmission.

\item A RDO algorithm using MAC-L beacons in order to achieve optimized video streaming under even lower complexity, as validated through experiments on a real-world AAV testbed and shown to minimize drops.
\item A co-design of a RDO algorithm with adaptive video encoding to achieve optimized video streaming, developed for both MPEG2 and MJPEG streaming and validated through experiments on a real-world AAV testbed, where it is shown to improve received video quality.
\end{itemize}


\section{Rate-Distortion Optimization Problem}
\label{sec:RDO}

The RDO problem aims to optimize the amount of distortion in a network against the rate. Distortion is defined as the degradation in media quality as packets are dropped. The rate represents the amount of
data, i.e. the number of packets transmitted per unit time. The data packets that comprise a stream vary in their importance in contributing to output quality and, conversely, distortion. As such, RDO is concerned with deciding which packets to drop based on media quality metrics, measuring both the deviation from the source material and the bit cost for each possible decision outcome. In other words, the problem aims to solve the question of, \emph{which} packets to select for transmission, \emph{when} to transmit them, and \emph{how} to transmit them (e.g., how many times), such that the expected distortion is minimized, subject to constraints on the expected rate.

In \cite{Chou:05}, the authors presented an algorithm that minimize the distortion $D$ for a given rate $R$. This is done by minimizing the Lagrangian $D+\lambda R$ for some Lagrange multiplier $\lambda$. This algorithm is based on off-line transmission policy computation and on-line transmission policy truncation. This problem is defined and solved for every data unit (i.e. video packet) $l$. As such, there exist a Lagrange multiplier $\lambda_l$ for every data unit $l$. Depending on the expected channel rate, 
the value $\lambda_l$ is a packet threshold used to decide if data unit $l$ is the optimal video packet for transmission at this time instant or not. Solving the rate distortion optimization problem is not efficient for embedded applications in a real-time setting as the transmission policy computation for every packet threshold $\lambda_l$ is time consuming and needs to be performed off-line. Instead, we propose novel RDO algorithms with low computational complexity in which the transmission policy is computed online in real time. In addition, solving the rate-distortion optimization problem presented in \cite{Chou:05} requires a mathematical channel model with parameters that are updated regularly. If used in a real-time system, this will add significant computational complexity. Instead of a mathematical channel model, we measure the channel state by using measurable physical quantities that are already available in default system operation and therefore will not require extra computation. We use two 
types of channel state feedback for our two proposed algorithms:
\begin{enumerate}
\item MAC-L ACKs\footnote{In 802.11 networks, a transmitting station can not listen for collisions while sending data, mainly because a station can not have its receiver on while transmitting a frame. As a result, the receiving station needs to send an acknowledgement (ACK) if it detects no errors in the received frame.}: In standard 802.11 wireless networks, the destination sends ACKs to the source when packets are successfully received. These ACKs are used in our algorithm at the source to measure the channel state. We call the low complexity RDO algorithm that uses MAC-L ACKs \textbf{LCRDO-Ack}.
\item MAC-L beacons\footnote{In 802.11 networks, access points periodically broadcast a beacon. The radio network interface card (NIC) receives these beacons while scanning and takes note of the corresponding signal strengths. The beacons contain information about the access point, including service set identifier (SSID), supported data rates, etc.}: In standard 802.11 wireless networks, stations send beacons. To further reduce complexitiy, these beacons can be used instead of more frequent ACKs to measure the channel state at the source. We call the low complexity RDO algorithm that uses MAC-L beacons \textbf{LCRDO-Beacon}.
\end{enumerate}
In addition to basic LCRDO variants, we co-design LCRDO-Beacon with adaptive video encoding algorithms using  both MPEG2 and MJPEG compressions. We call this third low complexity RDO algorithm with adaptive co-design \textbf{LCRDO-Adaptive}. 

The proposed LCRDO-Ack, LCRDO-Beacon, LCRDO-Adaptive algorithms have two important characteristics: 
\begin{enumerate}
\item Real-time compatibility, where the transmission policy/channel state is computed on-line.
\item Lower complexity in terms of computational processing requirements.
\end{enumerate}


\section{Testbed}\label{sec:testbed}
We demonstrate real-world performance of our optimized RDO algorithms in the context of Horus, a testbed composed of a network of AAVs communicating wirelessly in an ad-hoc fashion. In our experimental analysis, the AAV nodes are equipped with video cameras and are capable of streaming packetized media between them. Our network consists of a fixed number of nodes, that are placed in a prespecified topology. 
In this network, sources are streaming video data in real time to destinations. 
The routing path is given a priori and the topology of the network is fixed throughout the experiment, but nodes are mobile and moving in fixed circular paths. This continuous movement results in a time-varying nature of the wireless channels and reveals the effectiveness of the implemented algorithms. We implement rate-distortion optimized algorithms and measure the system performance for these networks.
  
\subsection{System Simulation}\label{sec:sys_sim}
We have setup a simulation of Horus network in the OMNeT++ network simulator
framework \cite{omnet} using the MiXiM package \cite{mixim}. To
accurately mimic and evaluate RDO behavior in the Horus setup, we
model both the time-varying nature of the network topology as well as
the modified protocol stack including our RDO layer on top of the
OMNET++ component library. Each AAV node is described using a standard OMNET++/MiXiM network
stack consisting of a physical layer, a MAC-L layer and an application
layer running the RDO optimized video streaming. Furthermore, we simulated mobile, time-varying network topologies using the circular motion module of OMNET++.

\subsection{System Implementation}\label{sec:sys_imp}

\begin{figure}[tb]
\centering
\includegraphics[width=1\columnwidth]{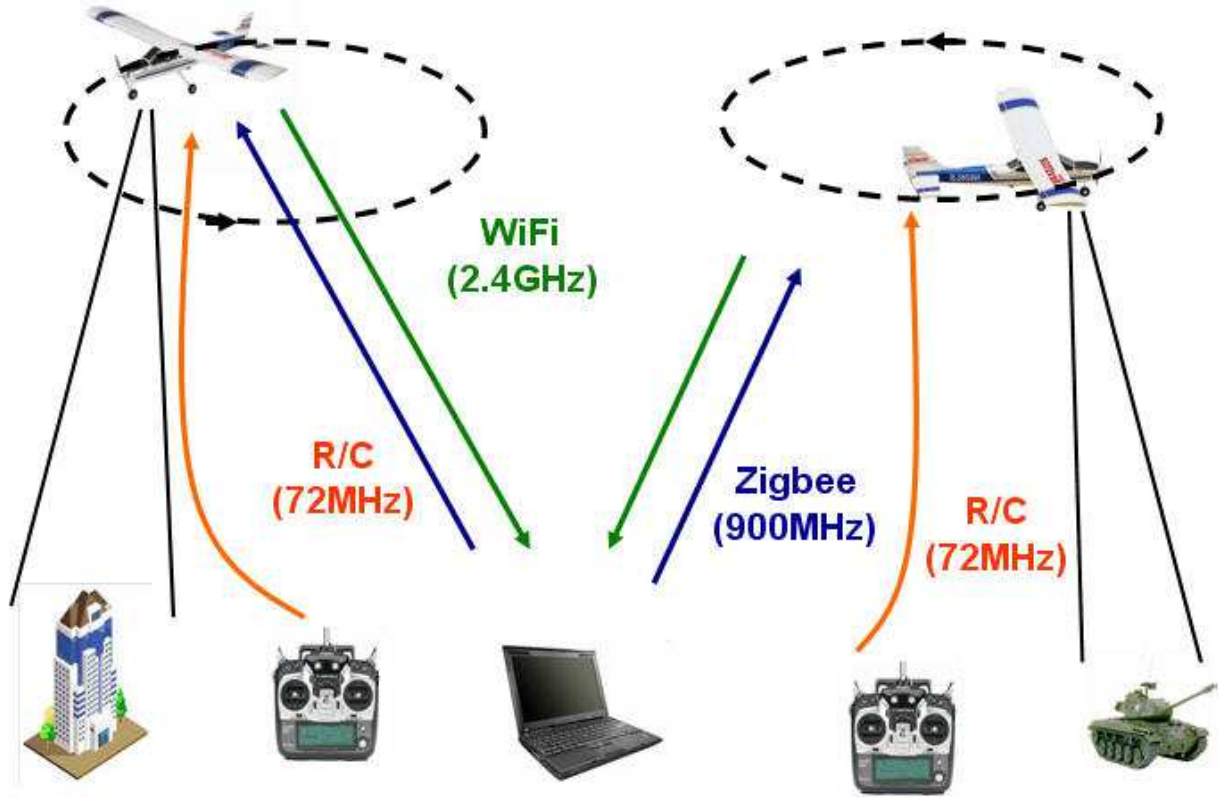}
      \caption{Horus project diagram.}
      \label{fig:project_diagram}
\end{figure}

\begin{figure}[tb]
\centering
\includegraphics[width=1\columnwidth]{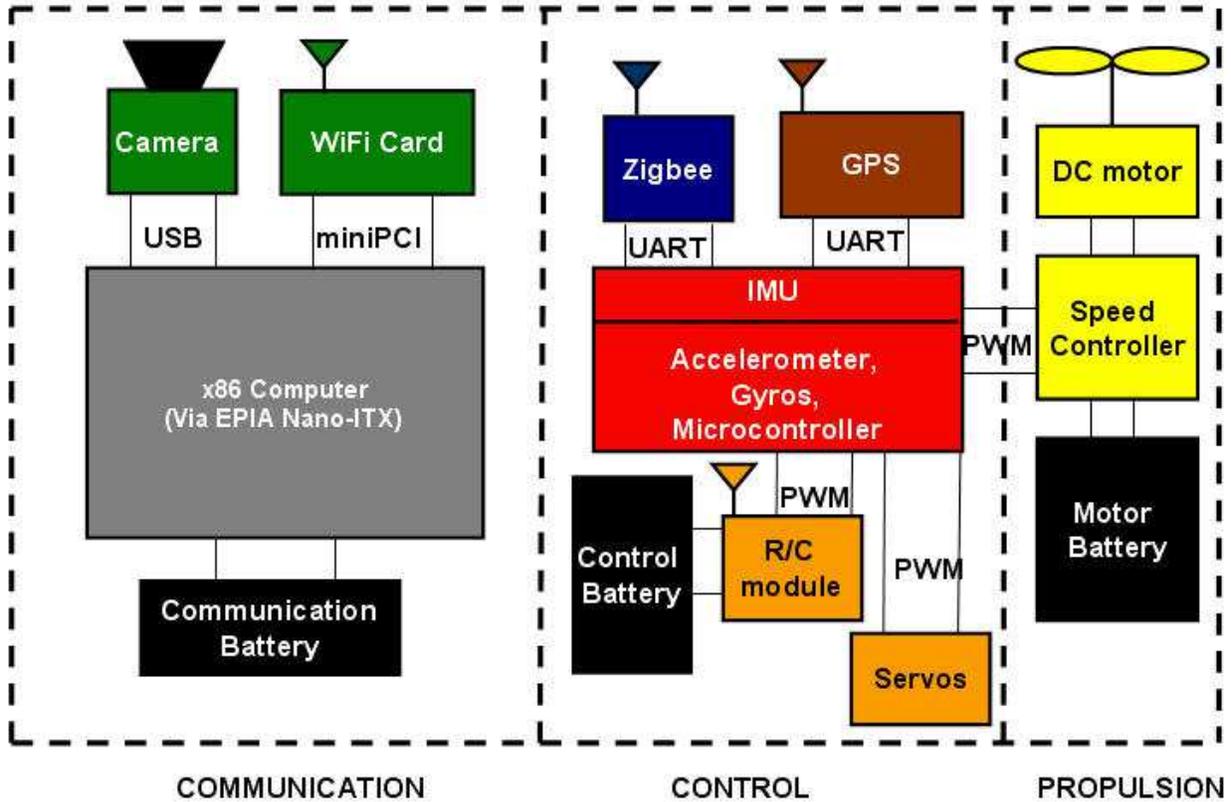}
      \caption{AAV block diagram.}
      \label{fig:block_diagram}
\end{figure}

In the following, we describe the actual implementation of our initial realization of the Horus network. We are considering AAVs as aerial nodes that form the network under test, see Figure \ref{fig:project_diagram}. The AAVs are controlled manually during take-off and landing using a remote control (R/C) module running at 72MHz. In the air, the AAVs are operating under automatic control by the on-board inertial measurement unit (IMU). The IMU uses stored way-points and GPS\footnote{GPS stands for Global Positioning System.} to follow a preprogrammed path. The path can be changed while the AAVs are in the air by uploading new waypoints from the ground laptop to the on-board IMU via a Zigbee radio link at 900MHz. 

Due to the nature of AAV nodes, there are constraints on the weight and dimensions of the components, as well as the power consumption, which directly affects the possible transmission range. We divide the AAV node architecture into three subsystems for propulsion, control, and communication, see Figure \ref{fig:block_diagram}\footnote{UART, PWM, and USB stands for Universal Asynchronous Receiver/Transmitter, Pulse-Width Modulation, and Universal Serial Bus, respectively. }.
The following lists represent the components that are used to construct our system nodes and their corresponding functions. The communication subsystem consists of the following components:

\begin{enumerate}
\item \textbf{Via EPIA Nano-ITX \cite{x86computer}:} is a x86 computer, which is the central unit for managing and operating the communication between nodes.
\item \textbf{Atheros WiFi radio \cite{atheros}:} is a wireless card for the IEEE 802.11 2.4 GHz frequency band, which is used for wireless video transmission.
\item \textbf{Video Camera \cite{camera}:} captures videos for our measurements and could be used for location identification and object recognition in future extensions of Horus.
\item \textbf{Communication Battery:} provides an independent power source for the communication subsystem.
\end{enumerate}
The control section consists of the following components:
\begin{enumerate}
\item \textbf{Zigbee \cite{zigbee}:}  a radio for the IEEE 802.15 900 MHz frequency band used for receiving way-points for autonomous AAV navigation.
\item \textbf{IMU unit \cite{imu}:} controls the AAV movement during flight and sustains the required flight paths and mobile network topology.
\item \textbf{GPS unit \cite{gps}:} determines the location of the AAV for use by the autopilot in the IMU.
\item \textbf{R/C module \cite{radio}:} uses the 72 MHz frequency band and is the main manual ground control of the AAV. In Horus, it is used to control take-off and landing of the AAV.
\item \textbf{Servo motors \cite{servos}:} act on flaps and rudder for controlling the direction and orientation of the AAV.
\item \textbf{Control Battery:} independent power source for the control subsystem.
\end{enumerate}
Finally, the propulsion section consists of the following components:
\begin{enumerate}
\item \textbf{DC motor \cite{motor}:} is the main moving force of AAV and is connected to the propeller.
\item \textbf{Speed Controller \cite{speedcontroller}:} controls the speed of the DC motor by regulating the input current.
\item \textbf{Motor Battery \cite{battery}:} main power source for AAV propulsion.
\end{enumerate}

\section{Low Complexity RDO with ACKs (LCRDO-Ack)}\label{sec:rdo1}
\begin{figure}[tb]
  \centering
    \includegraphics[width=1\columnwidth]{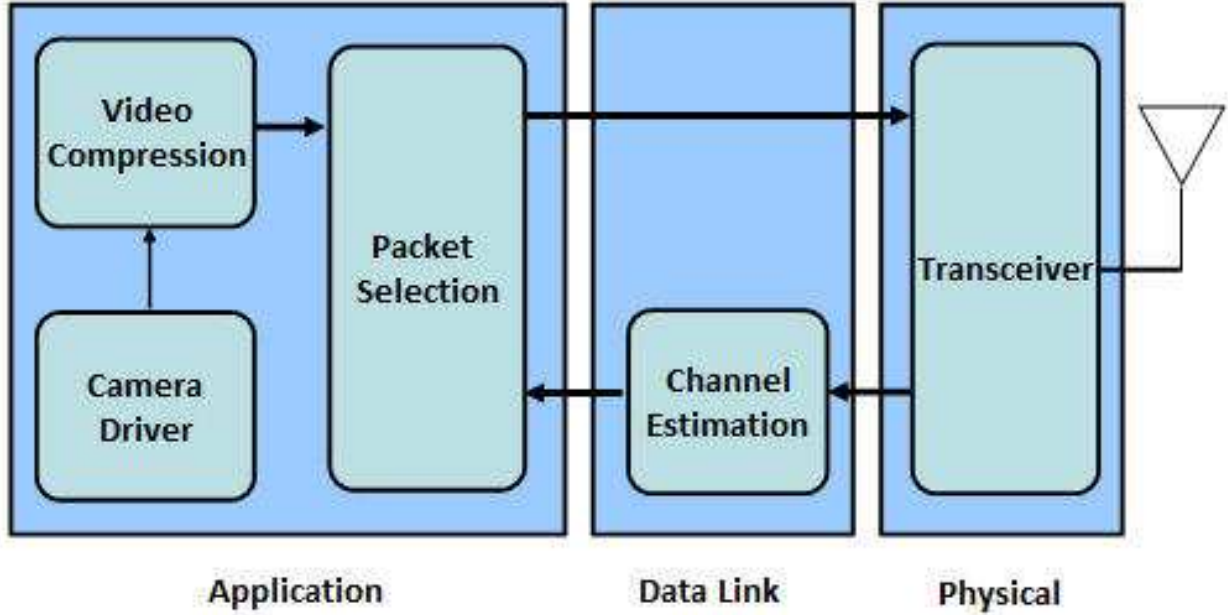}
      \caption{Conceptual block diagram of the LCRDO-Ack algorithm.}
      \label{fig:rate_distortion}
\end{figure}

We realize the LCRDO-Ack algorithm as part of the application layer of our network. A conceptual block diagram for the LCRDO-Ack algorithm mapped to OSI network layers is shown in Figure \ref{fig:rate_distortion}.  The video streams received from the camera are compressed into frames with different priorities.  The \emph{channel estimator} estimates the channel condition based on the received ACKs from previously sent packets. The \emph{channel estimator} block is by default receiving ACKs from the destination in IEEE 802.11 networks. The \emph{packet selector} block is responsible for selecting the suitable packet for transmission based on (1) the information received from the channel estimator and (2) the packet timestamp associated with each packet. In our case, the decision for retransmitting a packet or sending a new packet is made by the \textit{packet selector} block according to the algorithm described in section \ref{sec:LCRDO-ACK}.

\subsection{LCRDO-Ack Algorithm}\label{sec:LCRDO-ACK}

In MPEG4/MPEG2 video encoding, frames are compressed with different ratios and dependencies, giving each frame a different priority. Frames are divided into packets, where the amount of data and hence the number of packets typically increases with the frame priority.  On the receiver side, these packets are recombined to form a frame that is then decoded. Therefore, losing a frame with high priority will lead to more deterioration in the quality of decoded video.

In line with existing video standards, we assume that the compressed
video is composed of three types of frames: frames with first priority
$f_1$ (also known as \emph{i frames}), frames with second priority
$f_2$ (\emph{p frames}), and frames with third priority $f_3$
(\emph{b frames}). For compression purposes, the video frames are
divided into Groups of Frames (GOF). Each GOF contains one $f_1$ frame
and a fixed number of $f_2$ and $f_3$ frames. The LCRDO-Ack
algorithm that we use to optimize the video transmission is
implemented in the application layer and is shown in Algorithm \ref{alg}.

        
        
\begin{algorithm}
\caption{LCRDO-Ack Algorithm}
\label{alg}
\begin{algorithmic}
\LOOP
\IF {$t < t_{GOF\_timestamp}$} %
	\IF {$ID_{frame}=1$} 
		\IF {$ID_{pkt} > ID_{finalf_1pkt}$}
        	\STATE {$ID_{frame}++$} \COMMENT{switch to next frame}
        \ELSE
        	\STATE {transmit $f_1$ packet}
        	\IF {success} 
        	\STATE{$ID_{pkt}++$} \COMMENT{switch to next $f_1$ packet}
        	\ENDIF
        \ENDIF
	\ELSIF {$ID_{pkt} \leq ID_{finalGOFpkt}$}
                \STATE {transmit $f_2$ and $f_3$ packets without retransmission}
                \STATE{$ID_{pkt}++$} \COMMENT{switch to next packet}
	\ENDIF
	
\ELSE
\STATE {$t_{GOF\_timestamp}+=\Delta t_{GOF}$}
\COMMENT{start new GOF interval}
\STATE {$ID_{frame} \gets 1$} \COMMENT {$ID_{frame}=1$ is dedicated for i frame in GOF}
\STATE {$ID_{pkt} \gets ID_{finalGOFpkt} + 1$} \COMMENT {packet ID set to first packet in GOF}
\ENDIF 
\ENDLOOP
\end{algorithmic}
\end{algorithm}

The algorithm consists of two main sections, \textit{timestamp check}
and \textit{packet selection}.  In the timestamp check, the algorithm
starts by comparing the current time with the timestamp of the current GOF,
$t_{GOF\_timestamp}$. If the GOF timestamp is not yet reached, packet
selection is performed. Otherwise, the algorithm aborts the current
GOF and advances to the next one. In packet selection (i.e.\ within the
same GOF), the transmitter first sends $f_1$ frame packets. Each $f_1$
packet is (re-)transmitted until it is successfully sent and the
algorithm can switch to the next one.  After finishing the
transmission of all $f_1$ packets, $f_2$ and $f_3$ packets are
sent. These lower priority packets are transmitted without waiting for
feedback. This avoid wasting time in packet retransmissions,
acknowledgements and reception times for these less important
packets. At the end of packet selection, the packet ID is incremented
to send the next packet until all packets in the current GOF are
transmitted or a timeout is reached.

The LCRDO-Ack algorithm is implemented on top of standard protocol stacks. As an additional optimization we can, however, modify the 802.11 MAC-L to further improve overall real-time performance. Specifically, when the MAC-L receives a packet with a $f_1$ flag, it uses its default behavior to retransmit the packet up to 3 times until an ACK is received. If no ACK is received after 3 tries, the MAC-L reports a drop to the application layer. By contrast, in case of $f_2$ or $f_3$ packets, the RDO algorithm does not require feedback about transmission success and a modified MAC-L can transmit the packets only once without waiting for any ACK. This
further reduces overall overhead and latencies.

\subsection{Distortion Measure}\label{sec:sim_dist}

We investigate the LCRDO-Ack algorithm using a multiplicative distortion
measure. The multiplicative distortion $D_m$ is initially set to the
maximum distortion level $D_0$. When frames are successfully received,
the distortion decreases by the number of frames of type $i$,
$N_{f_i}$, multiplied by all the frames of higher priority
successfully received within a GOF.  This gives zero weight if frames
of higher priority have not been successfully received.
$D_m$ is evaluated every GOF as
\[
D_m =D_0 -N_{f_1}\Big(1+N_{f_2}(1 + N_{f_3})\Big); \:\:0 \leq D_m \leq D_0.
\]
We also define the sum of multiplicative distortion $D_M$ as the sum
of $D_m$ for all transmitted GOFs:
\[
D_M=\sum_{GOF}{D_m}.
\]
$D_M$ is the distortion measure we use to compare different
transmission protocols.

\subsection{Experimental Setup}

For our experiments, we assume a topology in which two or three AAVs
(hosts) move in fixed circular patterns with a radius of 40 meters and
a distance of 380 meters between the circles centers, see Figure \ref{fig:project_diagram}.  Nodes send data packets in a one-hop fashion
over a 802.11 wireless connection, where the source node transmits
packets directly to the destination node. Next to the transmission
under test, we include a third node that simultaneously transmits
other packets not related to the main video stream.  This setting
allows us to analyze RDO transmission in the presences of high
interference and consequently when experiencing a large packet loss.

We simulated this setup in OMNET++ using a
simple path loss channel model as the one most closely resembling
AAV-to-AAV conditions with little to no fading and no shadowing
effects. The continuous motion of the nodes in circular paths leads to
time-varying channel effects and a network packet drop rate that
depends on the relative position of the AAVs.  We use different path
loss exponents $\alpha$ to model and experiment with normal and worst
case channel conditions. For worst-case analysis, we assume an
exponent $\alpha$ of 3.7. Together with interference from a third node
as described above, we observe an overall packet drop rate of 40\%,
which allows for comparison of various transmitters under realistic
conditions.

For testing the LCRDO-Ack algorithm, we compare it against conventional
transmission algorithms. Overall, we define three types of
transmitters:
\begin{enumerate}
\item \textit{Transmitter without ACK:} transmits every packet without waiting for an ACK from the receiver.
\item \textit{Transmitter with ACK:} retransmits every packet until it receives an ACK for each packet.
\item \textit{Transmitter running the LCRDO-Ack algorithm:} implements the LCRDO-Ack algorithm described previously.
\end{enumerate}

\subsection{Results}\label{sec:results}

To compare different transmitters, we run the network simulator for
each transmitter under the exact network conditions mentioned in the
previous subsection. We specify 3500 packets to be transmitted from
the source to the destination node. For simplicity, we fix the number of
packets per frame for a given priority frame. For the payload parameters, the number of priority $f_1$, $f_2$, and $f_3$ frames per GOF are 1, 2, and 6, respectively, and the number of packets per frame $f_1$, $f_2$, and $f_3$ are 50, 20, and 10, respectively.

Our performance investigation for this problem includes both a network
measure (e.g. number of packets drops) and an optimization measure
(e.g. quality of the received media). We illustrate both using an
easy-to-visualize proxy variable, which is a counter at the
receiver. This counter counts the number of successfully received
packets and is incremented until the end of the frame is reached, upon which the counter is reset to zero. Counting then starts in the same manner for the next frame, and so on. The counter can determine if the
received packet is in the current frame or not by checking the packet
ID number associated with it. This way, plotting the counter values
over time visualizes the performance of the receiver with
respect to the video frames. These values are later used to measure
the distortion.

The No-ACK Transmitter sends packets continuously without receiving
any ACKs from the receiver. This leads to loss of packets with equal
probability for different priority frames, which results in a 40\%
loss of the first priority packets essential to decode the other
packets sent within a GOF. Due to this behavior, about 40\% of frames are not completely received
leading to high distortion, for more details see Figures in \cite{Ahmed3}.

The ACK Transmitter sends new packets only after receiving an ACK for
the previous packet, and it otherwise continues to retransmit the same
packet. Therefore, all the frame packets within the current GOF
timestamp are received successfully regardless of their priority. The
frames received after their GOF timestamp are dropped, but at the same
time cause more delay to build up with time. This accumulative delay
is caused by retransmission and ACKing of packets of lower
priority. Due to this delay, the number of frames that are lost
increases as the transmission continues, causing a large degradation
in the video quality over time. This leads to a significant increase
in the distortion at the end of simulation. The number of frames lost per GOF increase as the
transmission continues, leading to high distortion $D_M$, see Figures in \cite{Ahmed3}. This transmitter experiences the highest
distortion when the simulation is allowed to run for a sufficiently
long time.

The LCRDO-Ack transmitter is designed to minimize the distortion measure and
give better performance than conventional transmitters. It retransmits
until an ACK is received only for the first priority packets.  Second
and third priority packets are sent without waiting for an ACK from
the destination. This guarantees that frames of priority $f_1$ will be
received at the receiver even under bad channel conditions, which is
the case in our simulation. The second and third priority frames,
$f_2$ and $f_3$ respectively, are dropped with about 40\% probability. Overall, the LCRDO-Ack 
transmitter guarantees a minimum video quality at the receiver side
and gives a better distortion than other transmitters.

In all simulated cases, the average packet drop rate is 40\%. In the
LCRDO-Ack case, it is guaranteed that most of these drops are lower priority
packets, which affects transmission quality less and makes the LCRDO-Ack algorithm more
robust.  In the No-ACK and ACK transmitters, these dropped packets can
be of any type of packet priority. Therefore, in these two cases, the
40\% drop rate significantly affects transmission quality.

\section{Low Complexity RDO with Beaconing and Adaptivity (LCRDO-Beacon and LCRDO-Adaptive)}
We implemented LCRDO-Beacon and LCRDO-Adaptive algorithms in our physical testbed setup as described in section \ref{sec:sys_imp}. On the Via EPIA computer mounted inside the AAV, we run a Ubuntu 10.10 Linux operating systems with kernel version 2.6.35. We use the GStreamer open source multimedia framework \cite{gstreamer} for video compression and the Click Modular Router open source network stack \cite{click} for implementing our transmission algorithm. The operating system used for the ground station (i.e.\ the ground laptop) is  Ubuntu 10.10 with Linux kernel 2.6.35. We implement the LCRDO-Beacon algorithm in connection with MPEG2 video compression and the LCRDO-Adaptive algorithm for both MPEG2 and MJPEG compressions.

\subsection{LCRDO-Beacon Algorithm}\label{sec:mpeg_rdo}

\begin{figure}[tb]
\centering
\includegraphics[width=1\columnwidth]{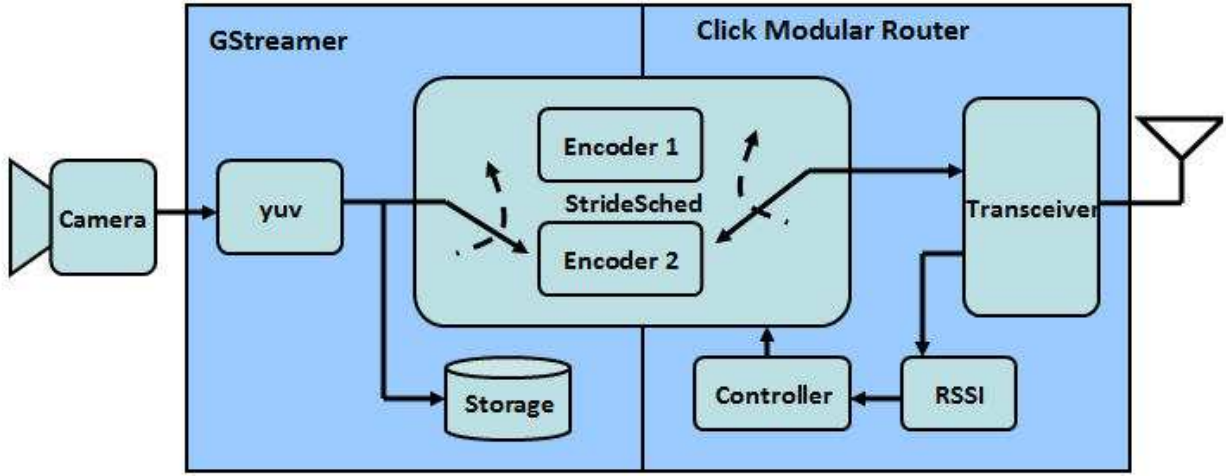}
      \caption{Universal block diagram of the implemented LCRDO algorithm at the transmitter.}
      \label{fig:rdo_mpeg_src}
\end{figure}

\begin{figure}[h!]
  \centering
    \includegraphics[width=0.5\columnwidth]{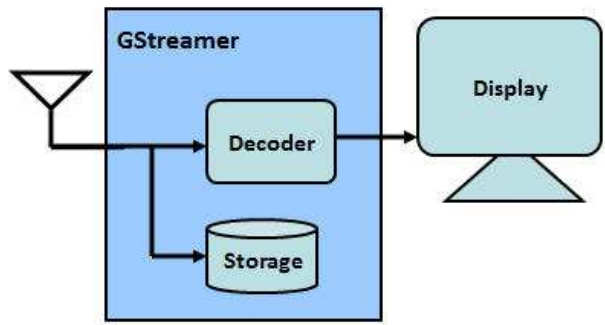}
      \caption{Universal block diagram of the implemented LCRDO algorithm at the receiver.}
      \label{fig:rdo_mpeg_dst}
\end{figure}

\begin{algorithm}
\caption{Implemented LCRDO Algorithm}
\label{alg:mpeg2RDO}
\begin{algorithmic}
\LOOP
\STATE {transmit packets received from the selected/active encoder}
\IF {transceiver receives a beacon} %
	\IF {$RSSI < X_1$} %
   		\STATE {Clear $Queue 1$ and $Queue 2$}   
   		\STATE {Switch to Encoder 2}
	\ELSE
		\IF {$RSSI > X_2$} %
  	 		\STATE {Clear $Queue 1$ and $Queue 2$}   
   			\STATE {Switch to Encoder 1}
		\ENDIF 
	\ENDIF 
\ENDIF
\ENDLOOP
\end{algorithmic}
\end{algorithm}

The LCRDO-Beacon algorithm, used in the system implementation, is computationally less complex than  the LCRDO-Ack algorithm (Algorithm \ref{alg}) used in the system simulation. The reason for further reducing the complexity of the algorithm are the high computational demands of MPEG2 video encoding, which requires most of the computation power of Via EPIA computer. The modifications  of switching from ACKs to beacons for channel state measurement ensure a more reliable real-time performance and concurrently show significant improvement in the general performance with respect to unoptimized methods, as will be shown in section \ref{sec:imp_results}.

The universal block diagram of the implemented LCRDO algorithm at the transmitter is show in Figure \ref{fig:rdo_mpeg_src}. For LCRDO-Beacon algorithm at the transmitter, the camera driver outputs raw video, which is resized to 160x120 (without loss of generality, this resolution is chosen to minimize the computation on the Via EPIA computer; the algorithm applies equally for higher resolutions). A stride scheduler chooses between different parameters for encoding the raw video into an MPEG2 stream. We realize two different encoders:
\begin{enumerate}
\item MPEG2 encoder with GOF\footnote{Group Of Frames (GOF) is the group of video frames that starts by an i frame then proceeded by p and b frames only. The MPEG video file is a sequence of GOFs; e.g. GOF=5 has one i frame and four p and b frames, while GOF=1 has only i frames and no p and b frames.}=5 and an overall frame rate of 25 frames per second (f/s). This corresponds to sending i, p and b frames with an i frame rate of 5 f/s.
\item MPEG2 encoder with GOF=1 and an overall frame rate of 5 f/s. This corresponds to sending i frames only at a rate of 5 f/s.
\end{enumerate}

Note that overall, this setup is equivalent to a general RDO architecture as described in section \ref{sec:rdo1} (Figure \ref{fig:rate_distortion}), where packets of p and b frames of a single, fixed MPEG2 encoder (running at 25 f/s) are selectively dropped depending on the chosen transmission policy. An equivalent implementation that alternates between two separate encoders as controlled by a stride scheduler was chosen due to limitations of the GStreamer-internal architecture.

The two outputs of the two MPEG2 compression and packet selection blocks are fed into the transmitter via two queues, $Queue 1$ and $Queue 2$. These queues are not drawn to simplify the block diagram. For later comparison purposes, a high quality reference copy of the original video is stored in compressed MPEG2 form (with GOF=5 and 25 f/s) using GStreamer. Inside the Click Modular Router, a transceiver realizes wireless transmission of encoded videos and wireless reception of signal strength beacons. The beacons received in the transceiver are passed through a Received Signal Strength Indicator (RSSI\footnote{The Atheros based card returns an RSSI value of 0 to 127 (0x7f) with 128 (0x80) indicating an invalid value. There is no specified relationship of any particular physical parameter to the RSSI reading.}) block that decodes the RSSI value and passes it on to a controller block. Finally, the controller block determines packet selection and controls the stride scheduler by executing Algorithm \ref{alg:
mpeg2RDO}. 

The implemented LCRDO algorithm is shown in Algorithm \ref{alg:mpeg2RDO}. In the LCRDO-Beacon algorithm case, the controller switches to the Encoder 2 (i frames only) in the video compression block whenever the RSSI drops below a value X1. Likewise, if the RSSI rises above a value X2, the controller switches to select the Encoder 1 (i, p and b frames) in the video compression block. The graph of the RSSI on the x-axis and the active encoder on the y-axis form a sharp hysteresis loop (figure removed due to figures limit). In the switching instant, all the contents of $Queue1$ and $Queue2$ are cleared. This is done to avoid sending any residual packets in the queue when switching back and forth between the different encoding modes, ensuring reliable real-time performance. 

The universal block diagram of the implemented LCRDO algorithm at the receiver is show in Figure \ref{fig:rdo_mpeg_dst}. For LCRDO-Beacon algorithm at the receiver, the received signal is decoded using a standard MPEG2 decoder and displayed on the monitor of the ground station (laptop). At the same time, the received video is stored in compressed MPEG2 format for later analysis.

\subsection{LCRDO-Adaptive Algorithm for MPEG2}\label{sec:mpeg_adp}

In addition to selectively dropping packets, a generalized method for performing RDO and improving the received video quality is to co-design RDO-type packet selection with adaptive video encoding. In such an approach, the video encoding rate is adjusted to the transmission rate in a distortion-optimized way. In addition to improving video quality, adapting encoding parameters to rate variations can significantly reduce average computational requirements in the real-time video encoder. Similar to the LCRDO-Beacon algorithm, such an adaptive approach has an operating mechanism in which the video encoder switches between different modes. Both algorithms require predetermined thresholds, i.e.\ $X_1$ and $X_2$, used in a sharp hysteresis loop. In the LCRDO-Beacon algorithm, the channel state measurement determines when to transmit both independent and dependent or when to drop dependent and only transmit independent frames. By contrast, for the adaptive algorithm, the channel state measurements determine when to 
transmit video at high quality, i.e. with high bit rate, and when to transmit video at low quality, i.e. with lower bit rate. Both algorithms require channel state measurements and seek to minimize distortion and maximize video quality.

The block diagram of the low complexity RDO with adaptive co-design (LCRDO-Adaptive) algorithm for MPEG2 transmissions is similar to the block diagram of LCRDO-Beacon with the exception that the stride scheduler allows choosing between two different MPEG2 encoders for video compression:
\begin{enumerate}
\item MPEG2 encoder with 5 frames per GOF, a frame rate of 25 f/s, and unlimited bit rate.
\item MPEG2 encoder with 5 frames per GOF, 25 f/s frame rate, and 100 kbps transmission rate.
\end{enumerate}


The block diagram of the receiver is similar to the LCRDO-Beacon receiver.

\subsection{LCRDO-Adaptive Algorithm for MJPEG/SMOKE}\label{sec:mjpeg}

The attractive property of Motion JPEG (MJPEG) video compression is the very low computational complexity compared to MPEG2 video compression. This comes at the expense of lower compression leading to higher bandwidth requirements. Furthermore, MJPEG compression is characteristic by all frames being independent, whereas MPEG2 compression has both independent and dependent frames. A variant of MJPEG compression is the SMOKE codec \cite{smoke}, which includes both JPEG frames and delta frames. JPEG frames are key-frame that are each followed by $N -1$ delta frames.  JPEG frames are independent while delta frames are constructed according to a motion estimation threshold using the key-frame. This threshold specifies how much each 16x16 block of pixels may differ before a new block is generated. A large value of the threshold causes more blocks to stay the same for more frames, decreasing bandwidth usage, but producing less accurate output. Likewise, a small number of delta frames between key-frames increase 
received video quality at the cost of a higher bit rate, and vise versa. 

The block diagram of the LCRDO-Adaptive for MJPEG/SMOKE transmitter is similar to the block diagram of LCRDO-Beacon with the following exceptions. The camera driver outputs raw video which is resized to 320x240 with frame rate of 10 f/s. A stride scheduler chooses between different parameters for encoding the raw video into an MJPEG/SMOKE stream. We realize two different encoders:

\begin{enumerate}
\item MJPEG/SMOKE encoder with high JPEG image quality of 80\% and $N$=8.
\item MJPEG/SMOKE encoder with low JPEG image quality of 30\% and $N$=8.
\end{enumerate}

The storage block stores high quality MJPEG/SMOKE compressed video (with JPEG image quality of 80\%, $N$=8 and 10 f/s) as reference for later analysis.

On the receiver side, see Figure \ref{fig:rdo_mpeg_dst}, the received signal is decoded using MJPEG/SMOKE decoder and displayed on the monitor of the ground station (laptop). At the same time, the received compressed video is stored as MJPEG/SMOKE format for later analysis.

\subsection{Distortion Measures}\label{sec:impDistortion}

The state of the art metric for comparative assessment of video quality is the MOtion-based Video Integrity Evaluation (MOVIE) Index \cite{movie}. However, since we experience frame drops in our video transmissions, it is difficult for us to apply the MOVIE index directly. Frame drops result in different video duration between the reference video stored in the transmitter and the video to be evaluated in the receiver. Therefore, we measure distortion in our experiments differently. First, we measure the number of frame drops experienced in the transmission, which represents the temporal loss in the received information. Second, we measure the spatial loss in the video by slicing the video into images/frames and comparing manually selected, representative best and worst received images/frames to their corresponding source images/frames.  

We qualify temporal distortion by comparing the difference in duration between the video viewed at the receiver and the high quality reference video at the transmitter. The high quality original video file is compressed by the encoder at the source while simultaneously being transmitted to the destination. At the destination, a copy of the video received is stored in compressed form while also being viewed in real-time. 

In addition, the MPEG2 decoder performs inter-frame estimation for the missing packets in the frame. The estimation process causes some of the frames to be distorted. To capture such frame distortions, we introduce an additional spatial metric. The spatial metric is intended to provide an approximate measure of the distortion in received frames when applied to the best and worst manually selected frames, as will be described in section \ref{sec:imp_results}. We measure the spatial distortion by applying the Structural SIMilarity (SSIM) \cite{ssim} index to the best and worst frames successfully received/reconstructed at the receiver. The SSIM index is a method for measuring the humanly perceived similarity between two images. It is a reference-based metric, where the assessment of image quality is based on a distortion-free reference image. We choose SSIM because it outperforms traditional methods, such as peak signal-to-noise ratio (PSNR) and mean squared error (MSE), which have proven to be inconsistent 
with human perception.



\subsection{Experimental Setup}\label{sec:exp_imp}

\begin{figure}[tb]
\centering
\includegraphics[width=3in]{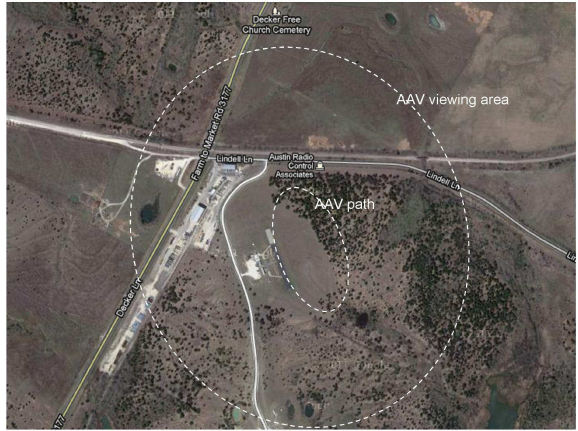}
      \caption{Map of the actual site used for running Horus experiments. The internal circle shows the AAV flight path. The outer circle shows the viewing area of the camera attached to the AAV.}
      \label{fig:map_view}
\end{figure}
In Horus, we consider two main network topologies:
\begin{enumerate}
\item Unicast network topology: one AAV moving in a circular pattern. The AAV is the source that records video of the landscape and transmits this video to the destination. The destination is a laptop in the ground station running Linux.
\item Multiple unicast network topology: two AAV moving in two separate circular paths. The AAVs are both sources that record video of the landscape and transmit them to one destination (as shown in Figure \ref{fig:project_diagram}). The destination is a laptop in the ground station running Linux.
\end{enumerate}

Note that a multicast network topology is easier to implement than a multiple unicast network topology. Two sources transmitting to one destination requires twice the bandwidth of a network in which one source transmitting to two destinations. Therefore, if a multiple unicast network topology is implementable and reliable using our proposed algorithms, then multicast network topology will almost certainly be implementable and reliable. 



We run the flight experiment at Lester Field \cite{lester} in Austin, Texas. In the flight experiments, the AAV records video of the landscape and transmits this video in real-time to the ground station (laptop). Throughout each flight, AAVs pass through four different flight phases, which correspond to different characteristics of the transmitted video:
\begin{enumerate}[A)]
\item The AAV is stationary on the ground and near the laptop (the ground station). This phase takes place during  AAV initialization and preparation.
\item The AAV is moving slowly and within close range of the laptop. This phase takes place when moving the AAV to the take-off runway.
\item The AAV is moving relatively fast and with moderate range of the laptop. This phase takes place at take-off and landing. In this phase, the recorded camera video is changing rapidly.
\item The AAV is moving fast and far away from the laptop. This phase takes place when the AAV is in the air.
\end{enumerate}

The AAV passes through these phases every time we run a flight experiment to test different transmission algorithms. In total, we conducted three different flight experiments:
\begin{enumerate}
\item Evaluating the performance of the LCRDO-Beacon algorithm with respect to unmodified MPEG2 video transmissions. We test the following:
  \begin{enumerate}
  		\item \textit{Unmodified transmitter with only i frames:} MPEG2 codec with GOF=1 and 5 f/s.
		\item \textit{Unmodified transmitter with all i, p and b frames:} MPEG2 codec with GOF=5 and 25 f/s. 
		\item \textit{Transmitter running the LCRDO-Beacon algorithm:} alternating MPEG2 codecs as discussed in section \ref{sec:mpeg_rdo}.
  \end{enumerate}
\item Evaluating the performance of the LCRDO-Adaptive algorithm for MPEG2 with respect to unmodified MPEG2 video transmissions. We test the following:
  \begin{enumerate}
  		\item \textit{Unmodified transmitter with 100 kbps:} MPEG2 codec with 100 kbps transmission rate and GOF=5 and 25 f/s.
		\item \textit{Unmodified transmitter with unlimited rate:} MPEG2 codec with unlimited rate and GOF=5 and 25 f/s.
		\item \textit{Transmitter running the LCRDO-Adaptive algorithm for MPEG2:} adaptive MPEG2 codec as discussed in section \ref{sec:mpeg_adp}.
  \end{enumerate}
\item Evaluating the performance of the LCRDO-Adaptive algorithm for MJPEG/SMOKE with respect to unmodified MJPEG/SMOKE video transmissions. We test the following:
  \begin{enumerate}
		\item \textit{Unmodified transmitter with low quality:} MJPEG /SMOKE codec with low quality of 30\% and $N$=8.
		 \item \textit{Unmodified transmitter with high quality:} MJPEG /SMOKE codec with high quality of 80\% and $N$=8.
		\item \textit{Transmitter running the LCRDO-Adaptive algorithm for MJPEG:} adaptive MJPEG/SMOKE codec as  discussed in section \ref{sec:mjpeg}.
  \end{enumerate}
\end{enumerate}

\begin{rem}
The hysteresis threshold values that we use in all the flight experiments are $X_1$=30 and $X_2$=50. 
\end{rem}

\subsection{Results}\label{sec:imp_results}

\begin{table}
\centering
\caption{Received video duration and average SSIM indices}\label{table:ImpResults}
\begin{tabular}{ | l | c | c |}
\hline
Experiment                         & Video duration   &  Average SSIM   \\ \hline
1(a) Send only i frames            & 0.68             & 0.773      	\\ \hline
1(b) Send all i, p, and b frames   & 0.53             & 0.770 		\\ \hline
1(c) LCRDO-Beacon                  & 0.76             & 0.777		\\ \hline
2(a) Send at 100kbps               & 0.71             & 0.717		\\ \hline
2(b) Send at unlimited rate        & 0.58             & 0.789 		\\ \hline
2(c) LCRDO-Adaptive  for MPEG2     & 0.76             & 0.762 		\\ \hline
3(a) Send at low quality           & 0.43             & 0.939		\\ \hline
3(b) Send at high quality          & 0.27             & 1				\\ \hline
3(c) LCRDO-Adaptive  for MJPEG     & 0.49           & 0.959 		\\ \hline
\end{tabular}
\end{table}
\begin{figure}[tb]
\centering
\includegraphics[width=1\columnwidth]{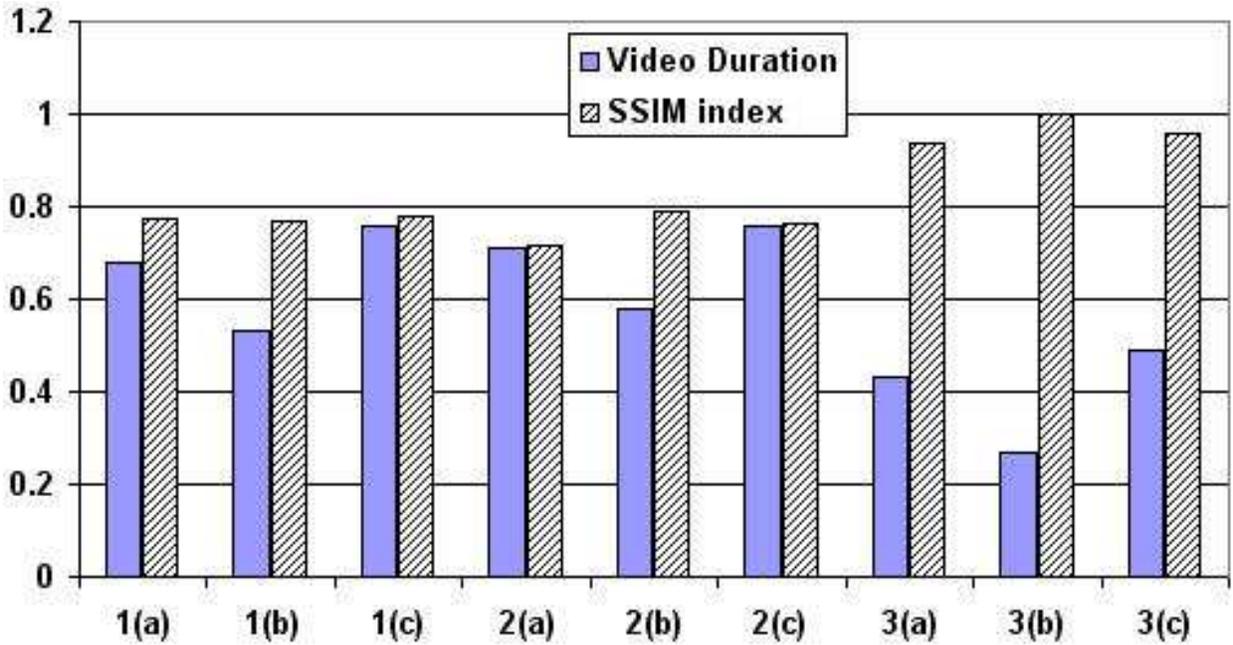}
      \caption{Received video duration and average SSIM indices.}
      \label{fig:ImpResults}
\end{figure}

For our experiments, we analyze both temporal and spatial distortion. For temporal distortion, we compare the video at the receiver/destination and the video at the transmitter/source with respect to video duration. For spatial distortion, we compare the SSIM index of 10 highest and 10 lowest spatially distorted video frames received at the destination. The frame selection is done manually by eye inspection. The number of frames selected from the flight phases A, B, C, and D are 2, 4, 4, and 10, respectively. The SSIM index is calculated for each frame with respect to the corresponding source frame. Then, the average, i.e. arithmetic mean, of SSIM indices for all the selected 20 frames is calculated for each case in each flight experiment.

Table \ref{table:ImpResults} and Figure \ref{fig:ImpResults} show the temporal distortion results for our experiments.  For every experiment, the fraction of the received video duration is highest for case (c).  Hence, we can conclude that a better optical flow of received video is observed when applying the LCRDO  algorithms compared to the other two cases for each experiment. Therefore, the temporal distortion observed by the user is minimal in the LCRDO algorithms. This corresponds to our experiences during real-time field observation.
 
\begin{rem}
When we run the same case in the same experiment under the same conditions multiple of times, we observe a 5\% difference in the results. Therefore, we can conclude that all the results presented here have an accuracy around 95\%.
\end{rem}
 



In terms of  spatial distortions, we observed that the average SSIM indices for case (a), (b), and (c) within each experiment are approximately equal, see Table \ref{table:ImpResults} and Figure \ref{fig:ImpResults}. We conclude that the average SSIM index is more dependent on the compression type compared to the transmission algorithm. Nevertheless, spatial distortions in the RDO case are always less than when streaming with a low quality encoder, approaching the level of high quality encoding yet with much better temporal behavior. The MJPEG/SMOKE codec outperforms MPEG2 codec in terms of its average SSIM index. This is due to inter-frame estimation performed in the MPEG2 decoder, which causes partially received frames to be viewed as distorted frames. By contrast, partially received frames are dropped in the MPJEG/SMOKE decoder, leading to higher temporal distortions instead. 

Finally, we conducted a flight experiment for the multiple unicast network topology mentioned in section \ref{sec:exp_imp}. The two AAVs transmit two different video signals simultaneously to a common ground station (laptop) using the LCRDO-Adaptive for MJPEG algorithm. The main observation is that we can display both videos with acceptable optical flow. Successful reception of multiple unicast flows should open more possibilities for future work in building larger AAV networks that can transmit/receive videos to/from multiple destinations/sources. As discussed previously, any implementation of the more complex multiple unicast case should easily transfer into a multicast environment. 

\section{Conclusion}
\label{sec:conclude}
Rate distortion optimization has its origins in information theory \cite{Cover_Thomas}. Rate distortion represents the minimum rate required to compress information at a distortion level of $D$. The rate distortion function with and without state is well known \cite{Cover_Thomas}, and can be computed for most cases algorithmically and in some cases in closed form. Our rate distortion algorithm thus aims at bridging theory and practice, by building an algorithmic framework for real-time rate distortion optimization that is low complexity and thus practically viable over an AAV testbed.


This paper constructs a reliable wireless networks test-bed for testing new wireless networks protocols called Horus. In this test-bed, we implement the problem of streaming packetized media over a wireless network using a rate-distortion optimized algorithm. We compare different transmission algorithms both in simulation and implementation. We give a comparative study to the design trade-offs to be considered to achieve a reliable and optimized video transmission. The main intuition that emerges from Horus is that in order to provide a good real-time video transmission performance, one should consider both the computation power and the bandwidth limitations. For low computation power, we find that MJPEG/SMOKE as the best choice. For moderate computation power, we have MPEG2 as the more suitable solution. For limited bandwidth but high computation power, MPEG4-H264 could potentially be the more suitable solution because it has better utilization of the available bandwidth compared to MJPEG/SMOKE and MPEG2. 
This is the trade-off between bandwidth and computation limitation and the video quality that can be achieved. 

\section{Future Work}\label{sec:future}

In this paper, we considered optimized video transmission. The current work on this topic is looking at extending this work to long term evolution (LTE) systems, say, optimal resource allocation in a cellular system. We are looking at the viability of the rate distortion optimization problem in joint optimization with resource allocation of video applications in cellular system, along the lines of \cite{Ahmed_Utility1, Ahmed_Utility2, Ahmed_Utility3, Haya_Utility1, Haya_Utility2},  to improve the quality of experience of mobile users. In addition, we plan on the inclusion of wireless communication capacity \cite{Ahmed1}, and interference alignment methods for MIMO communication systems \cite{Ahmed2}.

\bibliographystyle{plain}
\bibliography{Thesis,KUser,dof}
\end{document}